\documentclass{llncs}

\title{First-order Logic as a Constraint Programming Language}

\author{K. R. Apt\inst{1,2} and C. F. M. Vermeulen\inst{1}}
\institute{CWI, P.O. Box 94079, 1090 GB Amsterdam, the Netherlands\\
\and
University of Amsterdam, the Netherlands}

\usepackage{amsfonts}
\usepackage{latexsym}
\usepackage{alltt}

\newcommand{\tupel}[1]{\langle #1 \rangle} 
\newcommand{\verz}[1]{\{ #1\} }

\newcommand{\infer}[0]{\mbox{\it infer}}

\newcommand{\elim}[0]{\mbox{\it elim}} 
\newcommand{\solve}[0]{\mbox{\it solve}} 
\newcommand{\myerror}[0]{\mbox{\sc error}} 
\newcommand{\states}[0]{\mbox{\sc states}}

\newcommand{\weg}[0]{\mbox{\sc drop}}

\newcommand{\mysem}[1]{[\! [#1 ]\!]}
\newcommand{\mycons}[0]{\mbox{\it cons}}

\newcommand{\csp}[0]{{\cal C}}

\newcommand{\ES}{\mbox{$\emptyset$}}

\newcommand{\la}{\mbox{$\:\leftarrow\:$}}
\newcommand{\ra}{\mbox{$\:\rightarrow\:$}}

\newcommand{\A}{\mbox{$\ \wedge\ $}}

\newcommand{\Or}{\mbox{$\ \vee\ $}}

\newcommand{\te}{\mbox{$\exists$}}

\newcommand{\LL}{\mbox{$\ldots$}}

\newcommand{\newMS}[1]{\mbox{$[\![{#1}]\!]$}}

\newcommand{\B}[1]{\mbox{$[\![{#1}]\!]$}}       
\newcommand{\C}[1]{\mbox{$\{{#1}\}$}}           

\newcommand{\error}{{\bf error}}

\newcommand{\var}{{\it Var}}

\newcommand{\sem}[1]{{\mbox{$[\![{#1}]\!]$}}}

\newcommand{\szkew}[1]{\relax \setbox0=\hbox{\kern -24pt $\displaystyle#1$\kern 0pt }%
\box0}
{\catcode`\@=11 \global\let\ifjusthvtest@=\iffalse}

\newcounter{oldmycaption}

\newcommand{\p}[2]{\langle #1 \ ; \ #2 \rangle}

\newcommand{\almazero}{{\sf Alma-0}}
\begin{document}
\date{}
\maketitle

\begin{abstract}
  We provide a denotational semantics for first-order logic that
  captures the two-level view of the computation process typical for
  constraint programming.  At one level we have the usual program
  execution.  At the other level an automatic maintenance of the
  constraint store takes place.
  
  We prove that the resulting semantics is sound with respect to the
  truth definition. By instantiating it by specific forms of
  constraint management policies we obtain several sound evaluation
  policies of first-order formulas.  This semantics can also be used a
  basis for sound implementation of constraint maintenance in presence
  of block declarations and conditionals.

\end{abstract}

\section{Introduction}
\label{sec:introduction}

By the celebrated result of Turing first-order logic is undecidable.
In particular, the question of determining for an interpretation
whether a first-order formula is satisfiable and finding a satisfying
substitution if it is, is undecidable.  Still, for many formulas this
question can be answered in a straightforward way.  Take for instance
the following simple formula interpreted over the standard
interpretation for arithmetics:
\begin{equation}
  \label{formula1}
y < z \A y = 1 \A z = 2  
\end{equation}
It is easy to see that it is satisfied by the substitution
$\C{y/{\bf 1}, z/{\bf 2}}$.  Similarly, it is easy to see that the
formula
\begin{equation}
  \label{formula2}
\neg (x = 1) \A x = 0  
\end{equation}
is satisfied by the substitution $\C{x/{\bf 0}}$. 

The question is whether we can capture this concept of
``straightforwardness'' in a natural way. Our first attempt to answer
this question was given in Apt and Bezem \cite{AB99} by providing a
natural operational semantics for first-order logic which is
independent of the underlying interpretation for it.  It captures the
computation process as a search for a satisfying substitution for the
formula in question.  Because the problem of finding such a substitution
is in general undecidable, we introduced in it the possibility of a
partial answers in the form of a special $\error$ state indicating a
run-time error.  In Apt \cite{Apt00} we slightly extend this
approach by explaining how more general equalities can be handled and
formulate it in the form of a denotational semantics for first-order
logic.  Unfortunately, both semantics are too weak to deal properly
with formulas (\ref{formula1}) and (\ref{formula2}): for both of them
the $\error$ state is generated.

In this paper we try to overcome these limitations by providing a
computational interpretation of first-order logic in the spirit of
constraint programming. According to this view the computation process
takes place on two levels. At one level we have the usual program
execution.  At the other level, in the ``background'' inaccessible to
the user, an automatic maintenance of the constraint store takes
place.
The problem we tackle is undecidable, so we introduce the possibility
of partial answers. They are modeled now by a non-empty
constraint store or the $\error$ state.

The automatic maintenance of the constraint store is modeled by a
parametric \emph{infer} operation that acts on states.  The idea of
an abstract \emph{infer} operation is due to Jaffar and Maher
\cite{jaffar-survey}. Here we consider it in presence of arbitrary
first-order formulas.  Because of this generality we can obtain
various sound realizations of the constraint store management by
appropriately instantiating \emph{infer}.  The correctness of this
approach is formalized in the form of an appropriate soundness result.
To establish it we need to assume some properties of the \emph{infer}
operation. They are formulated as five ``healthiness'' conditions. 

To illustrate the benefits of this view of first-order logic and to
show the scope of the soundness result, we discuss several ways of
instantiating the $\infer$ parameter to specific constraint management
policies.  Examples include admission of a constraint store consisting
of arbitrary first-order formulas, restriction to a constraint store
consisting only of atomic constraints, and restriction to a constraint
store consisting only of arbitrary first-order positive formulas.  We
can also discuss in this framework in a uniform way unification, an
algorithm for solving equations and disequations over the Herbrand
algebra, and Gaussian elimination in presence of arithmetic
constraints.

On the more practical side, these considerations lead to specific
implementation proposals of the constraint store in presence of block
declarations and conditionals, here modeled, respectively, by means
of existential quantification and of negation, conjunction and
disjunction.

To clarify these issues we return to formula
(\ref{formula1}).  If we do not admit a constraint store, as in the
semantics of \cite{AB99} and \cite{Apt00}, its evaluation yields the
$\error$ state, since we cannot evaluate $y < z$ without knowing the
values for $y$ and $z$.  But if we do allow atomic constraints in the
store, we can postpone the evaluation of $y < z$ and the evaluation
yields the substitution $\C{y/{\bf 1}, z/{\bf 2}}$.

Next, let us reconsider formula (\ref{formula2}).  If only atomic
formulas are allowed as constraints, the evaluation of this formula
yields the $\error$ state, since we can neither evaluate $\neg (x =
1)$ nor add this formula to the constraint store. If, however, negated
formulas are allowed in the constraint store the substitution
$\C{x/{\bf 0}}$ is an answer.  The soundness theorem states that each
computed substitution satisfies the evaluated formula.

The question of providing an appropriate semantics for first-order
logic in the spirit of constraint programming could be approached by
taking for a formula $\phi(\bar{x})$ a clause $p(\bar{x}) \la
\phi(\bar{x})$, where $p$ is a new relational symbol and by applying
to it a transformation in the style of Lloyd and Topor \cite{LT84}.
The outcome would be a constraint logic program that uses negation.
After clarifying how to deal properly with negation this could yield a
rather indirect answer to the question we study. In contrast, our
approach, expressed in the form of a denotational semantics, is much
more direct and conceptually transparent: the meaning of each formula
is expressed directly in terms of the meaning of its constituents and
it is parametrized in a simple way by the \emph{infer} operation.

The rest of the paper is organized as follows.
In Section \ref{sec:general} we introduce the $\infer$ operation and
discuss in detail the requirements we impose on it. The main
difficulty has to do with the appropriate treatment of existential
variables.  In Section \ref{sec:sem} we define our denotational
semantics.  Next, in Section \ref{sec:apt} we show that the proposed
semantics subsumes the denotational semantics provided in
\cite{Apt00}.  Then in Section \ref{sec:specifi} we discuss various
increasingly powerful forms of constraint store management, each
modeled by means of a particular $\infer$ operation.
Finally, in Section \ref{sec:related} we discuss related work.

\section{Towards the denotational semantics}
\label{sec:general}

Below we work our way towards our proposal for the
denotational semantics in several steps,
first introducing the basic semantic ingredients, then discussing
the crucial conditions on the $\infer$ parameter and finally
presenting the denotational semantics for first order
logic with $\infer$ parameter.
In the next section we will then state the soundness result
for the semantics. The proof details are referred to the appendix.
We discuss several ways of instantiating the $\infer$ parameter
to show the scope of the soundness result.
In the final section we review the goals and results and look ahead
to further developments.

\paragraph{{\bf Preliminaries}}
\label{sub:prelim}

Let's assume that an algebra ${\cal J}$ is given over which we want to perform
computations.
The basic ingredient of the semantic universe will be the set of states,
$\states$. States come
in two kinds. First we have an $\myerror$ state, which remains unanalyzed.
All other states consist of two components: one component is a constraint 
store $\csp$, the other a substitution $\theta$. Such a state is then written 
$\tupel{\csp ;\theta}$. As always, a substitution 
$\theta$ is a mapping from variables to terms. It assigns a term 
$x\theta$ to each variable $x$, but there are only finitely many variables 
for which $x\neq x\theta$. These variables form $dom(\theta )$, the 
domain of $\theta$. 
The application of a substitution $\theta$ to a term $t$, written 
$t\theta$, is defined as usual. We denote the empty substitution by
$\epsilon$.

A constraint store $\csp$, 
is simply a finite set of formulas of first order logic. In many applications
there are extra requirements on the syntactic form of a constraint store,
but for now we keep things as general as possible. $\bot$ is a special
formula which is always false.

Throughout the paper we try to limit the number of brackets 
and braces
as much as possible. In particular, for a finite set $\verz{A_1,\ldots,A_n}$
we will often write $A_1, \ldots,A_n$. Also, 
we write $\infer\tupel{\csp ;\theta }$ instead  of 
$\infer(\tupel{\csp  ;\theta})$, etc.

\paragraph{{\bf The treatment of local variables: dropping things}}
\label{sub:local}

An important ingredient of the set up is
the ${\it DROP}_u$ mapping on states. It is the way we deal with
local variables. This works in two steps: first
we define the substitution {\it DROP}$_u(\theta )$ for each 
variable $u$ and substitution $\theta$, as in \cite{Apt00}:\\

\begin{tabular}{ll}
$u {\it DROP}_u(\theta )$ $=\; u$\\
$x {\it DROP}_u(\theta )$ $=\; x\theta$ \mbox{for all other variables $x$}\\
\end{tabular}\\

So, ${\it DROP}_u$ makes the current value of $u$ disappear,
thus capturing the idea of a local variable to the substitutions.
But we also have another component in states: the constraint
store. Dropping $u$ from such a set of formulas 
compares to existential quantification over $u$. 
There is one little extra point to take care of, however: in a state
$\tupel{\csp  ;\eta}$ the information that $\eta$ provides about the value
of $u$ is implicitly available  to $\csp$. Therefore, we perform the 
quantification $\exists u$ only
after adding the information about the value of $u$ explicitly to $\csp$.
Also the values $y\eta$ in which $u$ appears have to be kept in mind.
We take the conjunction of the equations $y=y\eta$ for all such
variables $y$ and write it as ${\bf y}={\bf y}\eta$. 
This leads us to the following formula that takes care of
the local variables in $\csp$. 

\begin{quote}
$\exists u\; (u=u\eta \;\wedge\; {\bf y}={\bf y}\eta\;\wedge\; \bigwedge\csp )$
\end{quote}

Note that this formula depends both
on $u$ and $\eta$. So, we cannot define a drop$_u$-mapping on constraint
stores alone: we have to know $\eta$ as well.

This formula expresses the information we are after in a uniform way,
but in `borderline cases' the syntactic format is awkward. For example,
if $\csp =\emptyset$, we get a trivial existential quantification
over the first two conjuncts. This existential quantifier is 
semantically harmless, but specific constraint propagation formalisms
simply do not work on existentially quantified formulas. Therefore we
rather adopt a format in which
the quantifier only appears if it is really necessary.

This is done in two steps:
first, the quantification over $u$ only matters for the
formulas in $\csp$ in which $u$ actually occurs. We make this explicit
in the definition by distinguishing  $\csp (u)$, the subset
of $\csp$ that contains exactly the formulas with the free variable 
$u$.
In the formula we use for the drop$_u$-mapping we can then always take
$\csp -\csp (u)$ outside the scope of the quantifier. This gives:
\begin{quote}
$\csp -\csp (u)\; ,\; \exists u\;
        (u=u\eta \;\wedge\; {\bf y}={\bf y}\eta\;\wedge\; \bigwedge\csp (u))$
\end{quote}
Finally, in the special case $\csp (u) =
\emptyset$, we leave out the existentially quantified formula altogether.

For the $\myerror$-state we simply set: 
$\mbox{\sc drop}_u\myerror =\myerror$.
To summarize, the  mapping $\mbox{\sc drop}_u$ is defined on states by 
the following cases:\\

\begin{tabular}{|llll|}
\hline
&&&\\
$\;\mbox{\sc drop}_u\tupel{\csp  ;\eta}$ & $=$&$ \tupel{\csp  ;
        \mbox{\it DROP}_u (\eta ) }$ & if $\csp (u)=\emptyset\;\;$\\[2mm]
$\;\mbox{\sc drop}_u\tupel{\csp  ;\eta}$ & $=$&
        $ \langle\;\exists u\; (u=u\eta \;\wedge\; {\bf y}={\bf y}\eta
        \;\wedge\; \bigwedge\csp (u)),$&\\
$\;$ & &        \hspace{2cm}$\; \csp -\csp (u);\;\;\;\;\;\;
                \mbox{\it DROP}_u (\eta )\rangle$ 
        & if $\csp (u)\neq\emptyset\;\;$\\[2mm]
$\;\mbox{\sc drop}_u \myerror $ & $=$ & $\myerror$ &\\
$\;$&&&\\
\hline
\end{tabular}

\paragraph{{\bf Conditions on infer}}
\label{sub:conditions}

Another important ingredient of the framework is the $\infer$ mapping.
$\infer$ maps a state to a set of states.
The $\infer$ mapping is the basic notion of computation in the semantics:
we do not specify what happens `within' the $\infer$ mapping.
This makes the set up extremely general: the $\infer$ steps can consist of 
calls to a constraint 
solver, like a unification algorithm or an algorithm for solving linear
equations over reals, calls to a constraint propagation algorithm,
or other atomic computation steps.
Several instances of
the $\infer$ mapping will be discussed in more detail later on.

We can almost get away with 
complete generality regarding $\infer$. To make sure that
the formalism respects first order logic, we have to make a few modest
requirements. 
Let us write $\tupel{\csp  ;\theta}\models_{\cal J}\phi$ for
$\csp\theta\models_{\cal J}\phi\theta$. In particular 
$\tupel{\emptyset ;\theta}
\models_{\cal J}\phi$ iff $\models_{\cal J}\phi\theta$.
Then the restrictions that we need in the soundness proof below,
read as follows:

\begin{description}
\item[(1) Equivalence:] 
        if $\tupel{\csp '  ;\theta '}\in \infer\tupel{\csp  ;\theta} $, then
        $\tupel{\csp  ;\theta}\models_{\cal J}\phi$ iff
        $\tupel{\csp ';\theta '}\models_{\cal J}\phi $ 
\item[(2) Renaming:] 
        if $\tupel{\csp '  ;\theta '}\in \infer\tupel{\csp  ;\theta} $, then
        also $\tupel{\csp '_v ;\theta '_v} \in \infer \tupel{\csp  ;\theta}$,

        where $\tupel{\csp '_v ;\theta '_v}$ is obtained from 
        $\tupel{\csp ' ;\theta '}$ by replacing all occurrences
        of $u$ by $v$ for a variable $u$ that is fresh w.r.t. 
        $\tupel{\csp  ;\theta}$ and a variable $v$ that is fresh w.r.t. both
        $\tupel{\csp  ;\theta}$ and $\tupel{\csp ' ;\theta '}$ 
\item[(3) Inconsistency:] if $\infer\tupel{\csp  ;\theta}=\emptyset$, then 
        $\tupel{\csp  ;\theta}\models_{\cal J}\bot$
\item[(4) Error:] $\infer\;\myerror =\verz{\myerror}$
\item[(5) Identity:] $\infer\tupel{\emptyset ;\theta}=
        \verz{\tupel{\emptyset ;\theta}}$
\end{description}

So, the $\infer$ mapping should respect logical equivalence,
i.e., the state $\tupel{\csp ' ;\theta '}$ that we reach starting 
from $\tupel{\csp  ;\theta}$, should still make the same formulas true.
Furthermore, the $\infer$ mapping should not be sensitive to the choice of
fresh variables: if $\infer$ works for $u$, 
it should also work for an alternative fresh variable $v$.
Finally, $\infer$ should respect falsity and the $\myerror$ state.\footnote{The
Identity requirement is not necessary for the proof of the soundness 
theorem, but it seems too natural to leave it out. Renaming is used only
in the proof of the Preservation/Persistence Lemma in
the case of the existential formula.}
When we talk about the consistency of states, we are dealing with
a three way distinction. We say that a state $\sigma$ is:
${\cal J}$-consistent,  if $\sigma \neq\myerror$ and 
        $\sigma \not\models_{\cal J}\bot$;
${\cal J}$-inconsistent, if $\sigma \neq\myerror$ and 
        $\sigma \models_{\cal J}\bot$;
error, if $\sigma =\myerror$.
For a set of states $\Sigma\subseteq\states$ we then distinguish:
$\mycons_{\cal J} (\Sigma )$ $=\; \verz{\sigma\in\Sigma:\; \sigma\mbox{ is
        $\cal J$-consistent} }$ and
$\mycons_{\cal J}^+ (\Sigma )$  $=\; \verz{\sigma\in\Sigma:\; \sigma\mbox{ is
        not $\cal J$-inconsistent} }$.
Usually it is clear to which ${\cal J}$ we refer and
we omit ${\cal J}$ from the notation.\\

\section{Denotational semantics}
\label{sec:sem}

We now define a denotational semantics for first order logic 
in which the $\infer$ mapping is a parameter. The parameter can be
set to give the semantics from Apt \cite{Apt00}, for example, but many 
other settings are available, as we will see below. This way
we obtain general results, that apply uniformly to various 
forms of constraint store management.

We define the mapping 
$\mysem{\phi}:\; \states\rightarrow\states$, using 
postfix notation.\footnote{We also sneak in the notation: 
$\Sigma\mysem{\phi}$ for 
$\bigcup_{\sigma\in\Sigma}\verz{\sigma\mysem{\phi}}$.}\\

\begin{tabular}{|ll|}
\hline
&\\ $\;\tupel{\csp  ;\theta}\mysem{A}\; $&
	$=\; \infer \tupel{\csp ,{A}  ;\theta}$
        \hspace{.8cm} for an atomic formula $A$
\\[2mm] $\;\tupel{\csp  ;\theta}\mysem{\phi_1\vee\phi_2}\;$&$ =\; 
        \tupel{\csp  ;\theta}\mysem{\phi_1}\; \cup\; 
        \tupel{\csp  ;\theta}\mysem{\phi_2}$
\\[2mm] $\;\tupel{\csp  ;\theta}\mysem{\phi_1\wedge\phi_2}\;$&$ =\;
        (\tupel{\csp ;\theta}\mysem{\phi_1})\mysem{\phi_2}$
\\[2mm] $\;\tupel{\csp  ;\theta}\mysem{\neg\phi}\;$&$ =
        \left\{
        \begin{tabular}{ll}
        $\infer\tupel{\csp  ;\theta}$ &
        \mbox{if } $\mycons^+ (\tupel{\csp  ;\theta}\mysem{\phi})=\emptyset$\\
        
        $\emptyset$ &
        \mbox{if }$\tupel{\csp ' ;\theta '}\in
                \mycons(\tupel{\csp  ;\theta}\mysem{\phi})$
        \mbox{ for}\\ &
        \mbox{some $\tupel{\csp ' ;\theta '}$ equivalent to
        $\tupel{\csp  ;\theta}$} \\[2mm]
        $\infer\tupel{\csp ,{\neg\phi}  ;\theta}$ &
        \mbox{otherwise}
        \end{tabular}
        \right.$\\[2mm]
$\;\tupel{\csp  ;\theta}\mysem{\exists x\; \phi}\;$&$ =\;
%
        \bigcup_{\sigma}\verz{\infer{\;\mbox{\sc drop}_u
                (\sigma )}}$

        where, for some fresh $u$,\\
        & \hspace{1.6cm} $\sigma$ ranges over $\mycons^+ (
        \tupel{\csp  ;\theta}\mysem{\phi\verz{x/u}})$ 
\\[2mm] $\;\myerror \mysem{\phi} \;$&$=\; \verz{\myerror}$  
		\mbox{ for all $\phi$}\\
&\\ \hline
\end{tabular}

\vspace{2mm}

The definition relies heavily on the notation that was introduced before. But
it is still quite easy to see what goes on. The atomic formulas are
handled by means of the 
$\infer$ mapping. Then, disjunction is interpreted as nondeterministic
choice,
and conjunction as sequential composition. For existential quantification
we use the $\weg _u$ mapping (for a fresh variable $u$). The $\myerror$ clause
says that there is no recovery from $\myerror$. In the case for negation,
three contingencies are present: first, the case where $\phi$ is 
inconsistent. Then we
continue with the input state $\tupel{\csp  ;\theta}$.
Secondly, the case where $\phi$ is already true in (a state 
equivalent to)
the input state. Then we conclude that $\neg\phi$ yields 
inconsistence, i.e., we get $\emptyset$. Finally, we add $\neg\phi$
to the constraint store $\csp$ if it is impossible at this point to reach
a decision about  the status of $\neg\phi$.

\label{sec:sound}

Next we show that the denotational
semantics with the $\infer$ parameter is sound. This 
amounts to two things:
{\it 1.}  successful
computations of $\phi$ result in states in which $\phi$ holds;
{\it 2.}  if no 
successful computation of $\phi$ exists, $\phi$ is false in 
the initial state.
\begin{theorem}[Soundness]
Let $\tupel{\csp ;\theta}$ and $\phi$ be given. Then we have:

\begin{enumerate}
\item If $\tupel{\csp' ;\theta '}\in\tupel{\csp ;\theta}\mysem{\phi}$, then
        $\tupel{\csp'  ;\theta '}\models_{\cal J}\phi$
\item If $\mycons^+ (\tupel{\csp ;\theta}\mysem{\phi})=\emptyset$ ,
        then $\tupel{\csp  ;\theta}\models_{\cal J}\neg\phi$.
\end{enumerate}
\end{theorem}

The proof of the theorem is by
a simultaneous induction on the structure of the formula $\phi$.
In the proof we need a preservation/persistence result, that we give
as a separate lemma.

\begin{lemma}[Preservation/Persistence]
\mbox{}

\begin{enumerate}
\item  If $\tupel{\csp  ;\theta}\models_{\cal J}\phi_1$ and 
$\tupel{\csp ' ;\theta '}\in \tupel{\csp  ;\theta}\mysem{\phi_2}$, then
$\tupel{\csp ' ;\theta '}\models_{\cal J}\phi_1$ \hfill{(validity)}
\item If $\csp \theta$ and $(\phi_1\wedge\phi_2)\theta$ 
are mutually consistent 
(in ${\cal J}$) and 

$\tupel{\csp ' ;\theta '}\in \mycons (\tupel{\csp  ;\theta}\mysem{\phi_2})$, 

then $\csp '\theta '$ and $(\phi_1\wedge\phi_2)\theta '$
are mutually consistent
(in ${\cal J}$).
\hfill{(consistency)}
\end{enumerate}
\end{lemma}

The lemma says that computations of $\mysem{\phi_2}$ will not disturb
the status of $\mysem{\phi_1}$: the computation preserves validity and 
consistency.
The proof of the lemma is by a 
simultaneous induction  on the structure of $\phi_2$.
Some proof details are given in the appendix. 
Here we continue by considering several instantiations of the general format.

\section{Modeling the denotational semantics of Apt \cite{Apt00}}
\label{sec:apt}

We start our analysis by recalling the semantics provided in
\cite{Apt00}.  The idea of this semantics is to provide a uniform
computational meaning for the first-order formulas independent of the
underlying interpretation and without a constraint store. This yields a
limited way of processing formulas in the sense that occasionally an
$\myerror$ may arise.  After we have reintroduced this semantics we shall
discuss a number of its extensions, all involving a specific
constraint store management.  So, let us recall the relevant
definitions.

\begin{definition} 
  Assume a language of terms $L$ and an algebra ${\cal J}$ for it.

\begin{itemize}
\item Consider a term of $L$ in which we replace some of the variables
by the elements of the domain $D$. We call the resulting object
a {\em generalized term}.

\item Given a generalized term $t$ we define its {\em ${\cal
      J}$-evaluation\/} as follows.  Each ground term of $s$ of $L$
  evaluates to a unique value in ${\cal J}$. Given a generalized term
  $t$ replace each maximal ground subterm of $t$ by its value in
  ${\cal J}$.  We call the resulting generalized term a {\em ${\cal
      J}$-term} and denote it by $\B{t}_{\cal J}$.
  
\item By a {\em ${\cal J}$-substitution\/} we mean a finite mapping
from variables to ${\cal J}$-terms which assigns to each variable $x$
in its domain a ${\cal J}$-term different from $x$.  We write it as
$\C{x_1/h_1,\dots,x_n/h_n}$.  
We define the notion of an application of a ${\cal J}$-substitution
$\theta$ to a generalized term $t$ in the standard way and denote it
by $t \theta$.  

\item A {\em composition of two ${\cal J}$-substitutions $\theta$
and $\eta$}, written as $\theta \eta$, is defined as the unique
${\cal J}$-substitution $\gamma$ such that for each variable $x$
\[
x \gamma = \B{(x \theta) \eta}_{\cal J}. 
\]
\end{itemize}
\end{definition}

The ${\cal J}$-substitutions generalize both the usual substitutions
and the valuations, which assign domain values to variables.
After these introductory definitions we recall the semantics
$\newMS{\cdot}$ of an equation between
two generalized terms (so {\it a fortiori}, between two terms).
Here and elsewhere we do not indicate the dependency of the semantics
on the underlying interpretation or algebra.

\begin{eqnarray*}
\begin{array}{lll}
\newMS{s=t}(\theta) & :=  &
\left\{
\begin{array}{ll}
\C{\theta \C{s\theta/\B{t\theta}_{\cal J}}} &  
\mbox{if $s\theta$ is a variable 
that does not occur in $t\theta$,} \\
\C{\theta \C{t\theta/\B{s\theta}_{\cal J}}} &  
\mbox{if $t\theta$ is a variable 
that does not occur in $s\theta$} \\
                                           &   
\mbox{and $s\theta$ is not a variable,} \\
\C{\theta} &  
\mbox{if $\B{s\theta}_{\cal J}$ and $\B{t\theta}_{\cal J}$ are identical,} \\
\ES &  \mbox{if $s\theta$ and $t\theta$ are ground and 
$\B{s\theta}_{\cal J} \neq \B{t\theta}_{\cal J}$,} \\
\C{\myerror} &  \mbox{otherwise.}
\end{array}
\right.
\end{array}
\end{eqnarray*}

Consider now an interpretation ${\cal I}$ based on an algebra ${\cal J}$.
Given an atomic formula $p(t_1, \LL, t_n)$  different from $s=t$ and a ${\cal
  J}$-substitution $\theta$ we denote by $p_{\cal I}$ the
interpretation of $p$ in ${\cal I}$.
We say that

\begin{itemize}

\item $p(t_1, \LL, t_n)\theta$ is {\em true\/} if
$p(t_1, \LL, t_n)\theta$ is ground
and $(\B{t_{1}\theta}_{\cal J}, \LL, \B{t_{n}\theta}_{\cal J}) \in p_{\cal I}$,

\item $p(t_1, \LL, t_n)\theta$ is {\em false\/} if
$p(t_1, \LL, t_n)\theta$ is ground
and $(\B{t_{1}\theta}_{\cal J}, \LL, \B{t_{n}\theta}_{\cal J}) 
	\not\in p_{\cal I}$.

\end{itemize}

To deal with the existential quantification we use the
$DROP_x$ operation defined in Section \ref{sub:local},
extended in the standard way to the
subsets of $Subs \cup \C{error}$.
Now $\newMS{\cdot}$ is defined by structural induction as follows.
$A$ is here an atomic formula different from $s=t$.

\begin{itemize}

\item $\newMS{A}(\theta) :=
\left\{
\begin{array}{ll}
\C{\theta} &  \mbox{if $A \theta$ is true,} \\
\ES &  \mbox{if $A \theta$ is false,} \\
\C{\myerror} &  \mbox{otherwise, that is if $A \theta$ is not ground,}
\end{array}
\right.
$

\item $\newMS{\phi_1 \A \phi_2}(\theta) 
	:=\newMS{\phi_2}(\newMS{\phi_1}(\theta))$,

\item $\newMS{\phi_1 \Or \phi_2}(\theta) 
	:=\newMS{\phi_1}(\theta) \cup \newMS{\phi_2}(\theta)$,

\item $\newMS{\neg \phi}(\theta) :=
\left\{
\begin{array}{ll}
\C{\theta} &  \mbox{if $\newMS{\phi}(\theta) = \ES$,} \\
\ES &  \mbox{if $\theta \in \newMS{\phi}(\theta)$,} \\
\C{\myerror} &  \mbox{otherwise,}
\end{array}
\right.
$
\item $\newMS{\te x \: \phi}(\theta) := DROP_{u}(\newMS{\phi\C{x/u}}(\theta))$,
where $u$ is a fresh variable.

\end{itemize}

The following example clarifies the way we interpret atoms and conjunction.
\begin{example} \label{exa:2a}
  
Assume the standard algebra for the language of arithmetic
with the set of integers as domain. 
We denote its elements by $\LL, \mathbf{-2,-1,0,1,2}, \LL$.
Each constant $i$ evaluates to the element {\bf i}.
We then have
\begin{enumerate}
\item 
$\newMS{y = z-1 \A z = x+2}(\C{x/{\bf 1}}) = 
\newMS{z = x+2}(\C{x/{\bf 1}, y/z-{\bf 1}}) =$

$ \C{\C{x/{\bf 1}, y/{\bf 2}, z/{\bf 3}}}, $

\item 
$ \newMS{y = 1 \A z = 1 \A y-1 = z-1 }(\varepsilon) = 
	\C{\C{y/{\bf 1}, z/{\bf 1}}},$

\item 
$\newMS{y = 1 \A z = 2 \A y < z }(\varepsilon) = \C{\C{y/{\bf 1}, z/{\bf 2}}},$

\item 
$\newMS{x = 0 \A \neg (x =1)}(\varepsilon) = \C{\verz{x/{\bf 0}}},$

\item \label{item1}
$\newMS{y-1 = z-1 \A y = 1 \A z = 1}(\varepsilon) = \C{\myerror},$

\item \label{item2}
$\newMS{y < z \A y = 1 \A z = 2}(\varepsilon) = \C{\myerror},$

\item \label{item3}
$\newMS{\neg (x = 1) \A x = 0}(\varepsilon) = \C{\myerror}.$
\end{enumerate}
\end{example}
So in this semantics the conjunction is not commutative and consequently it
is important
in which order the formulas are processed.
This semantics is a special case of the semantics provided in 
Section \ref{sec:sem}.
It is obtained by using the following $\infer$ relation:

\begin{itemize}
\item $\infer \p{A}{\theta} := \verz{\p{\emptyset}{\eta}:\;
        \eta \in\mysem{A}(\theta )}$ for an atomic formula $A$,
where we identify $\p{\ES}{\myerror}$ with $\myerror$,
\item $\infer \p{{\cal C}}{\theta} := \verz{ \myerror }$ 
for all other states $\p{{\cal C}}{\theta}$.
\end{itemize}

The relevant `embedding' theorem is the following one.
\begin{theorem}[Embedding]\mbox{}

\begin{itemize}
\item $\eta \in\mysem{\phi}(\theta )$ iff  
	$\p{\emptyset}{\eta} \in \p{\emptyset}{\theta}\mysem{\phi}$.

\item $\myerror \in\mysem{\phi}(\theta )$ iff  
$\myerror \in \p{\emptyset}{\eta}\mysem{\phi}$.
\end{itemize}
\end{theorem}

\section{Specific constraint store managements}
\label{sec:specifi}

\newcommand{\aux}[0]{\mbox{\it aux}}
\newcommand{\step}[0]{\mbox{\it step}}
\newcommand{\split}[0]{\mbox{\it split}}
\newcommand{\splitstep}[0]{\mbox{\it splitstep}}

We now illustrate the generality of our approach by presenting various
increasingly powerful forms of constraint store management. Each of
them is obtained by a particular propagation $\step$
that works on {\it special states} and is executed 
whenever and as-long-as it can be applied.
$\aux$ is our name for the maximal repetition of the $\step$.\footnote{Note 
that maximal 
repetition of one $\step$ is just one strategy for constraint management.
Already Jaffar and Maher \cite{jaffar-survey} mention other options, 
distinguishing
for example, {\it quick-checking}, {\it progressive} and {\it ideal} {\it CPL}
systems. Of course, our set up can also accommodate such variations.} 
So, 
$\aux$ is a procedure on special states that is the least fixed point of
$\aux = \step\circ\aux$.
Then we can define the
$\infer$ mapping as follows:

\begin{tabular}{ll}
$\infer\; \error$ & $=\; \verz{\error}$\\
$\infer \tupel{\emptyset ;\theta }$ & $=\; \verz{\tupel{\emptyset ;\theta}}$\\
$\infer \tupel{\csp ;\theta }$ & $=\; \aux\;\tupel{\csp ;\theta}\;\;\;$
                for a special state $\tupel{\csp ;\theta}$\\
$\infer \tupel{\csp ;\theta }$ & $=\; \verz{\error} \;\;\;\;\;$ otherwise\\
\end{tabular}

Now the examples are obtained by a specification of the special states and
the $\step$ procedure.
In each case it is then straightforward to check that
the adopted definition of $\infer$ satisfies the 
conditions we put on it in Section \ref{sub:conditions}.
Consequently, in each case the Soundness Theorem holds.  Informally,
in each case we provide a sound constraint store management.

\paragraph{{\bf Equations as active constraints}}
\label{subsec:equations}

Below, following Jaffar and Maher \cite{jaffar-survey}, we make a
distinction between \emph{active} and \emph{passive} constraints.  In
our framework active constraints are the ones that are capable of
changing the values of the variables, while the passive ones boil down
to formulas that become tests after an appropriate instantiation.

As an example how active constraints can be modeled using the
presented semantics consider unification as a way of
solving equality constraints.  To model it we choose as the
underlying algebra the Herbrand algebra, the universe of which
consists of the set of all ground terms of the language $L$.

The constraint stores of special states only contain 
equations.  The equations are active, and each step 
consists of unification, whenever possible. So, we put:

\[
\step \p{\emptyset}{\theta}:= \emptyset
\]
\[
\step \p{{\cal C}, s = t}{\theta}
 :=
\left\{
\begin{array}{ll}
\C{\p{{\cal C}}{\theta\eta}} &  
	\mbox{if $\eta$ is an mgu of $s\theta$ and $t\theta$,} \\
\ES                 &    \mbox{if $s\theta$ and $t\theta$ are not unifiable.}
\end{array}
\right.
\]

Other specific forms of active constraints can be modeled
in our framework in an equally straightforward way.

\paragraph{{\bf Atoms as passive constraints}}

\label{subsec:atoms}

The drawback of the semantics defined in the previous
section is that it yields $\error$ when
a wrong order of conjuncts is accidentally chosen. A possible remedy is to use
atoms as passive constraints, i.e., to move the atoms that currently 
evaluate to $\error$ to the constraint store instead. 

For the handling of passive constraints we include a $\split$
procedure on special states to isolate the passive constraints:
$\split\tupel{\csp ;\theta}\; =\; \tupel{\csp_p,\csp_a ;\theta}$, where
$\csp_p$ is a list of the constraints that are passive when evaluated 
by $\theta$ and 
$\csp_a$ is a list of the constraints that are active when evaluated 
by $\theta$.
When this is done, we perform a $\step$ on the  active constraints.
Next we re-group the constraints to reconsider the
active-passive $\split$ in the new state. So, the step
we perform in the auxiliary procedure 
is a composed action: $\aux = \splitstep \circ \splitstep$
and  $\splitstep =\split\circ\step$.\footnote{We 
ignore various implementation details regarding the particular
choice of an active
constraint and the distinction between lists and sets.}

In the current example we set the $\split$ procedure as
indicated: we regard the atoms that would evaluate to
$\error$ passive. Then the $\step$ works as follows:\\

\begin{tabular}{lll}
$\step \tupel{\csp_p ;\theta}$
        &  $= \; \verz{\tupel{\csp_p ;\theta}}$
        & if no active constraints occur\\
$\step \tupel{\csp_p,\csp_a, s=t ;\theta}$
        & $=\; \verz{\tupel{\csp_p,\csp_a
                 ;\theta\eta}}$
        & if $\eta$ is an mgu of $s\theta ,\; t\theta$\\
$\step \tupel{\csp_p,\csp_a, s=t ;\theta}$
        & $=\;\emptyset$
        & if $s\theta ,\; t\theta$ cannot be unified\\
$\step \tupel{\csp_p,\csp_a, A ;\theta}$
        & $=\;\verz{\tupel{\csp_p,\csp_a
                 ;\theta}}$
        & if $A\theta$ is true\\
$\step \tupel{\csp_p,\csp_a, A ;\theta}$
        & $=\; \emptyset $
        & if $A\theta$ is false\\
\end{tabular}\\

Then the $\splitstep := \split\circ\step$ combines the two actions 
and $\aux$ 
repeats the $\splitstep$ until no more active constraints are left
to remove.
Reconsider now the formulas from items (\ref{item1}) and
  (\ref{item2}) of Example \ref{exa:2a}.  We now have
\begin{quote}$
\p{\ES}{\varepsilon}\sem{y-1 = z-1 \A y = 1 \A z = 1} = $

$
\p{y-1 = z-1}{\varepsilon}\sem{y = 1 \A z = 1} = $
$ \C{\p{\ES}{\C{y/{\bf 1}, z/{\bf 1}}}}$
\end{quote}
and
\begin{quote}
$
\p{\ES}{\varepsilon}\sem{y < z \A y = 1 \A z = 2} = 
\p{y < z}{\varepsilon}\sem{y = 1 \A z = 2} = $

$
\C{\p{\ES}{\C{y/{\bf 1}, z/{\bf 2}}}}.$
\end{quote}
This shows the difference brought in by this
$\infer$ procedure.
However, in the case of the formula from item (\ref{item3}), we still have
\[
\p{\ES}{\varepsilon}
\sem{\neg (x = 1) \A x = 0} = \C{\error}.
\]

\paragraph{{\bf Equations as active and passive constraints}}
\label{subsec:ac-pas}

In general, equations can be both active and passive constraints.  For
example, linear equations over reals can be active and non-linear ones
passive.  To model computation in their presence we choose as the
underlying algebra the standard algebra for the language of arithmetic
with the set of real numbers as the domain.
The special states are the ones that just have equations
in the constraint store.
Next, we use a $\split$ procedure that regards the linear equations
as active and the  non-linear ones as passive.
Using standard arithmetic operations each linear equation can be
rewritten into one of the following forms:\\

\begin{tabular}{ll}
$\bullet$ & $0 = 0$, \\

$\bullet$ & $r = 0$, where $r$ is a non-zero real, and \\

$\bullet$ & $x = u$, where $x\in \var$ and $u$ a linear
expression not containing $x$.  \\
\end{tabular}\\

This leads to the following
definition of the propagation $\step$:
\begin{quote}$
\step  \p{{\csp_p,\emptyset}}{\theta} := \{\p{{\csp_p}}{\theta}\}
$\end{quote}
\begin{quote}$
\step  \p{\csp_p,{\cal C}_a, s = t}{\theta}
 :=
\left\{
\begin{array}{ll}
\p{\csp_p,{\cal C}_a}{\theta} &  
        \mbox{$s\theta = t\theta$ rewrites to $0 = 0$,} \\
\ES                                      &    
        \mbox{$s\theta = t\theta$ rewrites to $r = 0$,}\\
                                         &    
        \mbox{where $r$ is a non-zero real,} \\
\p{\csp_p,{\cal C}_a}{\theta \C{x/u}} &  
        \mbox{$s\theta = t\theta$ rewrites to $x = u$,} \\
\end{array}
\right.
$\end{quote}
The last clause models in effect the Gaussian elimination step, now in
presence of linear and non-linear equations.

\paragraph{{\bf Negative literals as passive constraints}}

\label{subsec:literals}

The $\infer$ methods introduced above allowed only atoms in the
constraint store of special states, that is to say an occurrence of non-atomic
formulas in the constraint store leads to an immediate error.  Let us
extend the $\infer$ method to allow for negative literals in the 
constraint store of special states. Now we can easily modify the 
definitions from "Atoms as passive constraints":
we regard states with finite sets of literals as special states
and regard the literals that would evaluate to
$\error$ as passive. Then the definition of the $\step$ is obtained
by having a literal $L$ instead of an atom $A$.
Now, in the case of 
the formula from item (\ref{item3}) of Example \ref{exa:2a}
we have
\begin{quote}
$
\p{\ES}{\varepsilon}\sem{\neg (x = 1) \A x = 0} = 
\p{\neg (x = 1)}{\varepsilon}\sem{x = 0} = 
$

$\step \p{\neg (x = 1)}{\C{x/{\bf 0}}} = 
\C{\p{\ES}{\C{x/{\bf 0}}}}.
$
\end{quote}

\paragraph{{\bf Equality and disequality constraints}}

We continue the previous example
for the case of
an arbitrary language of terms together
with equality and disequality constraints.\footnote{We ignore
the notational distinction between the disequation $s\neq t$ and the
negation $\neg (s=t)$ for the moment.} 
We adapt the definition by
having as active constraints all equations as well as those
disequations that are ground or of the form $t\theta\neq t\theta$.
The $\split$ of $\tupel{\csp ; \theta}$ now produces 
$\tupel{\csp_p,\csp_a ;\theta}$ with 
$\csp_p \; =\; s_1\neq t_1,\ldots , s_n\neq t_n$, a list of all the 
disequations $s_i\neq t_i\in\csp$ for which
$(s_i\neq t_i)\theta$ is not ground and not of the form $t\neq t$. 
The definition of the $\step$ then is:\\

\begin{tabular}{lll}
$\step \tupel{\csp_p ;\theta}$
        &  $= \; \verz{\tupel{\csp_p ;\theta}}$
        & if no active constraints occur\\
$\step \tupel{\csp_p,\csp_a, s=t ;\theta}$
        & $=\; \verz{\tupel{\csp_p,\csp_a
                 ;\theta\eta}}$
        & if $\eta$ is an mgu of $s\theta ,\; t\theta$\\
$\step \tupel{\csp_p,\csp_a, s=t ;\theta}$
        & $=\;\emptyset$
        & if $s\theta ,\; t\theta$ cannot be unified\\
$\step \tupel{\csp_p,\csp_a, s\neq t ;\theta}$
        & $=\;\verz{\tupel{\csp_p,\csp_a
                 ;\theta}}$
        & if $s\theta\neq t\theta$ is true\\
$\step \tupel{\csp_p,\csp_a, s\neq t ;\theta}$
        & $=\; \emptyset $
        & if $s\theta \neq t\theta$ is false\\
\end{tabular}\\

Then we get, for example

\begin{quote}
$
\p{\ES}{\varepsilon}\sem{f(x) \neq f(y) \A g(x,b) = g(a,y)} = $

$\step \p{ f(x) \neq f(y)}{\C{x/a, y/b}} = 
\C{\p{\ES}{\C{x/a, y/b}}}$.
\end{quote}

In general, if no $\error$ occurs, we can expect 
$\tupel{\emptyset ,\; \epsilon}\sem{\phi}$
to contain special states from which all active constraints are removed,
i.e., states of the form 
$\tupel{\csp ; \theta}$ where $\csp$ is a list of inequations $s\neq t$
such that
$s\theta \neq t\theta$ is passive. It follows from the
independence of inequations of \cite{Col84} that over an infinite
Herbrand Universe such a constraint store is consistent, i.e., has a grounding
solution $\eta$. For such an $\eta$ we can then conclude: 
$\models s_i\theta \neq t_i\theta$ 
(for each $1\leq i\leq n$) and $\models \phi\theta\eta$.

The grounding solution $\eta$ can not be built up during the computation of 
$\sem{\phi}$. This is clear from the example $x\neq y\; \wedge\; x=c$. If
we make the choice $\verz{x/d ,x/c}$ as a grounding solution for 
$x\neq y$ too soon,
we are no longer able to deal with $x=c$ later on. Hence we can benefit
from the independence of inequations only after the computation of 
$\sem{ x\neq y\; \wedge\; x=c}$ has been completed.

\paragraph{{\bf Existential formulas as passive constraints}}

At this point only literals are allowed in the constraint store.
We can easily extend the current store management to one
in which also existential formulas are allowed in the constraint store.
To this end we need some quantifier elimination procedure
$\elim$ that is able to deal with at least
some form of existential quantification. 
Then we can have $\step := \elim$.

\paragraph{{\bf Arbitrary formulas as passive constraints}}

The previous constraint store management can be extended by allowing 
arbitrary formulas in the constraint store. 
This makes sense as soon as we have some decision procedure $\solve$
that is able to deal with at least some type of negative formulas.
Then we can have $\step := \solve$.

\section{Rationale and Related Work}

\label{sec:related}

As clarified in Section \ref{sec:apt} the soundness result established
here generalizes the appropriate result provided in \cite{Apt00}. The
drawback of this semantics was that it yielded error as answer for
several clearly satisfiable formulas, like the ones considered in the
introduction.

Our interest in a semantics that models constraint management in a
sound way stems from our attempts to add constraints to the
programming language \almazero{} of Apt et al.  \cite{ABPS98a}.
\almazero{} extends imperative programming by features that support
declarative programming. This language allows us to interpret the
formulas of first-order logic (without universal quantification) as
executable programs. In Apt and Schaerf \cite{AS99a} we proposed to
extend \almazero{} by constraints but found that this led to situations
in which the customary interpretation of the conditionals by means of the
implication is unsound.

Using the above considerations we can provide a simple sound
interpretation of the \texttt{IF B THEN S ELSE T END} statement.
Namely, it is sufficient to interpret it in logic as $(B \A S) \Or
(\neg B \A T)$, written in the \almazero{} syntax as \texttt{EITHER B;
  S ORELSE (NOT B); T END}.  This interpretation requires that
negative literals, here \texttt{NOT B}, are used as passive
constraints. On the implementation level backtracking is then needed
but the above interpretation can be reduced to the customary
implementation of \texttt{IF B THEN S ELSE T END} if the condition
\texttt{B} evaluates to {\bf true} or {\bf false} irrespectively of
the constraint store.

As already mentioned in the introduction the modelling of the
constraint store maintenance by means of an abstract $\infer$
mechanism is due to Jaffar and Maher \cite{jaffar-survey}. In their
framework the computation mechanism of constraint logic programming is
modeled, so local variables (modeled by existential variables) and
negation are absent but recursion is considered. Additionally, only
conjunctions of atomic formulas are allowed as constraints.

In \cite{JMMS98} several semantics for constraint logic programming
are compared. In this paper a mapping {\it solv} is used that allows for
inconsistency checks during the computation. {\it solv} can vary with
the intended application, just like our {\it infer} parameter, but,
unlike {\it infer}, it cannot model arbitrary constraint propagation
steps. In fact, in \cite{JMMS98} the constraint propagation steps take
place only at the end of each the computation.

An alternative approach to model the essentials of constraint
programming is provided by the concurrent constraint programming (ccp)
approach pioneered by Saraswat \cite{saraswat-ccp-93} and Saraswat,
Rinard and Panangaden \cite{saraswat-semantic}. In this scheme the programs can
also be considered as formulas with the difference that the atomic
\textbf{tell} and \textbf{ask} operations are present and that the
parallel composition connective is present.  The idea captured by this
model is that the processes interact by means of a constraint system
using the \textbf{tell} and \textbf{ask} operations.  The constraint
system is a set of constraints equipped with the entailment operation.

The ccp programs can be written in a logical way by dropping the
``\textbf{tell}'' context around a constraint and by interpreting the
\textbf{ask}($c$) statement as the implication $c \ra$.  However, in
spite of this logical view of ccp programs it is not clear how to
interpret them as first-order formulas with the customary semantics.
In Fages, Ruet and Soliman \cite{FRS01} a logical semantics of ccp
programs is given by interpreting them in intuitionistic linear logic.
Both the denotational semantics for this language and the correctness
(in the assertional style) of ccp programs were considered in a
number of papers, see, e.g., de Boer et al.  \cite{dBdPP95} and de Boer
et al. \cite{dBGMP97}.  How to add to this framework in a
  sound way negation was studied in Palamidessi, de Boer and Pierro
  \cite{dBPdP97}.  By the nature of this approach the study of the
  constraint store management captured here by means of the $\infer$
  mechanism is absent in this framework.

\section*{Acknowledgement}
We thank Catuscia Palamidessi for helpful discussion on the subject 
of ccp programs.

\section*{\bf Appendix: the proofs}
\label{appie}

In this appendix we give proof details of the Soundness Theorem
and Preservation/Persistence Lemma. 
Both proofs are by 
simultaneous inductions on the structure of the formula. 
%
We focus on the existential quantification cases, which are the most subtle.
We use the notation  $\models_{\cal J}\phi[\overline{a}]$
to indicate the assignment of values $\overline{a}$ to (at least) 
the free variables in $\phi$. In the case of the lemma it will be convenient
to standardize this as follows: we are concerned with $\csp\theta$ and
$\phi_i\theta$ (for $i=1,2$) and denote the values for the
free variables shared by the $\phi_i\theta$ by $\overline{d}$. Then we use  
the values
$\overline{c}$ for the remaining free variables in $\phi_1\theta$ and 
$\overline{e}$
for the remaining free variables in $\phi_2\theta$. Finally, we denote the 
values of the remaining free variables in $\csp\theta$ by $\overline{b}$.
So, we will mostly use blocks of the form
$[\overline{b},\overline{c},\overline{d}, \overline{e}]$.\\

\noindent{\bf Proof of Preservation/Persistence Lemma 1}:

\noindent{$\circ$ \bf{atoms}}: 
In the atomic case $\phi_2 = A$ for some atom $A$ and 
$\tupel{\csp '' ;\theta ''}\in\infer \tupel{\csp ,{A} ;\theta}$.
Straightforward application of property (1) does the trick.

%
%
%
%
%
%
%

\noindent{$\circ$ \bf{disjunction}}:
In this case $\phi_2 = (\psi_1\vee\psi_2)$ and $\tupel{\csp '' ;\theta ''}\in
\tupel{\csp ;\theta}\mysem{\psi_i}$ for some $i=1,2$. 
In this situation the inductive hypotheses apply straightforwardly.

%
%
%
%
%
%
%

\noindent{$\circ$ \bf{conjunction}}:
In this case $\phi_2=(\psi_1\wedge\psi_2)$ and $\tupel{\csp '';\theta ''}\in
\tupel{\csp ';\theta '}\mysem{\psi_2}$ for some 

\noindent$\tupel{\csp ';\theta '}\in \tupel{\csp ;\theta}\mysem{\psi_1}$. 
Now two applications of the inductive hypothesis are required.
For preservation of validity this is straightforward. For preservation
of consistency it works as follows:
%
by assumption
$\models_{\cal J} (\csp\theta ,(\phi_1\wedge (\psi_1\wedge\psi_2 ))\theta )
[\overline{b},\overline{c},\overline{d},\overline{e}]$.
So,
$\models_{\cal J} (\csp\theta ,(\phi_1\wedge \psi_1)\theta )
[\overline{b},\overline{c_1},\overline{d_1},\overline{e_1}]$,
restricting the $\overline{d}$ and $\overline{e}$ to the relevant variables
and moving some of the $\overline{d}$ values to $\overline{c_1}$.
By induction hypothesis we get
$\models_{\cal J} (\csp '\theta ',(\phi_1\wedge \psi_1)\theta ')
[\overline{b_1'},\overline{c_1'},\overline{d_1'},\overline{e_1'}]$.
Next the induction hypothesis (for $\phi_1\wedge\psi_1$ and $\psi_2$)
provides
$\models_{\cal J} (\csp ''\theta '',
((\phi_1\wedge \psi_1)\wedge\psi_2)\theta '')
[\overline{b''},\overline{c''},\overline{d''},\overline{e''}]$,
as required.

\noindent{$\circ$ \bf{negation}}:
In this case $\phi_2=\neg \psi$. We have to distinguish cases

\begin{itemize}
\item $\phi_2$ is 'true': $\tupel{\csp '';\theta ''}\in
\infer \tupel{\csp ;\theta}$ (and 
$\mycons^+ (\tupel{\csp ;\theta}\mysem{\psi})=\emptyset$. )
Now property (1) gives the results.
\item $\phi_2$ is 'false': in this case $\tupel{\csp ;
\theta}\mysem{\phi_2}=\emptyset$ and both (i) and (ii) are void.
\item 'otherwise': we get $\tupel{\csp '';\theta ''}\in 
\infer\tupel{\csp ,{\neg\psi};\theta}$.
The results follow from first order logic and property (1).
\end{itemize}


\noindent{$\circ$ \bf{existential quantification}}:
Now $\phi_2=\exists x\; \psi$ for some $x$, $\psi$. So, consider
$\infer \;\mbox{\sc drop}_u\tupel{\csp ';\eta}$
for some consistent 
$\tupel{\csp ';\eta}\in\tupel{\csp ;\theta}\mysem{\psi\verz{x/u}}$ ($u$ fresh).
By property (1) it suffices to consider
$\mbox{\sc drop}_u\tupel{\csp ';\eta}=$
$\langle (\csp '-\csp '(u,{\bf y})) ,\;\exists u\; (u=u\eta\wedge {\bf y}=
{\bf y}\eta \wedge \csp '(u,{\bf y})) 
;\;  \mbox{\it DROP}_{u}(\eta ) \rangle$.
We call ${\it DROP}_{u,{\bf y}}(\eta )=\theta ''$.
\begin{enumerate}
\item By assumption $\tupel{\csp ;\theta}\models_{\cal J}\phi_1$.
By induction hypothesis 
$\tupel{\csp ' ;\eta}\models_{\cal J} \phi_1$.
From this 
$(u=u\eta\wedge \csp ')\mbox{\it DROP}_u(\eta )\models_{\cal J} (u=u\eta\wedge
\phi_1)\mbox{\it DROP}_u(\eta)$.
So,
$(u=u\eta\wedge \csp ')\mbox{\it DROP}_u(\eta )\models_{\cal J} 
\phi_1\mbox{\it DROP}_u(\eta)$.
Repeating this for the ${\bf y}$, we get
$(u=u\eta\wedge{\bf y}={\bf y}\eta\wedge \csp ')\theta '' \models_{\cal J} 
\phi_1\theta ''$.
By (2) we may assume that this holds for some $u$ that does 
not occur in $\phi_1\theta ''$.
Hence the implicit overall universal quantification over $u$ can be
replaced by an existential quantification over $u$ on the left hand side of
the sequent. 
This gives
\begin{quote}
$((\csp '-\csp '(u,{\bf y}))\; \wedge\; 
\exists u\; (u=u\eta\wedge{\bf y}={\bf y}\eta\wedge 
\csp '(u,{\bf y})))\theta '' \models_{\cal J} 
\phi_1\theta ''$.
\end{quote}
Now we can safely re-instantiate the values of the variables in
{\bf y} to obtain the same for 
${\it DROP}_u(\eta)\; =\; \theta ''\cup\verz{{\bf y}/{\bf y}\eta}$
\begin{quote}
$((\csp ' -\csp '(u,{\bf y}))\; \wedge\; 
\exists u\; (u=u\eta\wedge{\bf y}={\bf y}\eta\wedge 
\csp '(u,{\bf y})))\mbox{\it DROP}_u(\eta ) \models_{\cal J} 
\phi_1\mbox{\it DROP}_u(\eta )$.
\end{quote}
\item
The assumption gives
$\models_{\cal J}(\csp\theta,\; (\phi_1\wedge\exists x\;\psi))\theta )
[\overline{b},\overline{c},\overline{d},\overline{e}]$.
From the induction hypothesis we obtain
$\models_{\cal J}(\csp '\eta,\; (\phi_1\;\wedge\;\psi\verz{x/u})\eta )
[\overline{b'},\overline{c'},\overline{d'},\overline{e'},f]$
where $f$ is the value of $u$. From this we conclude
$\models_{\cal J}(\csp '\eta,\; (\phi_1\;\wedge\;\exists x\; \psi)\eta )
[\overline{b'},\overline{c'},\overline{d'},\overline{e'},f]$.
So, we can be sure that suitable values for all the variables in $\csp '$ and
$(\phi_1\wedge\exists x\;\psi )$ are available, if we use the values in
the block 
$[\overline{b'},\overline{c'},\overline{d'},\overline{e'},f]$
and the substitution $\eta$ as a middle man.
But then we can also assign these values directly to the variable 
$u$, eliminating the middle man $\eta$. This way we get 
values 
$[\overline{b''},\overline{c''},\overline{d''},\overline{e''}]$
such that
$\models_{\cal J}
(((\csp '-\csp '(u,{\bf y}))\wedge 
u=u\eta\wedge {\bf y}={\bf y}\eta\wedge \csp 
'(u,{\bf y}))\mbox{\it DROP}_u(\eta ),$
\hfill{$(\phi_1\wedge\exists x\;\psi)\mbox{\it DROP}_u (\eta ))
[\overline{b''},\overline{c''},\overline{d''},\overline{e''}]$.}
From this the consistency of $\mbox{\sc drop}_u \tupel{\csp ';\eta}$ and
$(\phi_1\wedge\exists x\;\psi )\mbox{\it DROP}_u(\eta )$ is clear.\hfill{$\Box$}
\end{enumerate}


\noindent{\bf Proof of Soundness Theorem 1}:

\noindent{$\circ$ \bf{atoms}}:
In case $\phi$ is an atomic formula $A$, $\tupel{\csp ;\theta}\mysem{\phi}
=\infer\;\tupel{\csp ,{A} ;\theta}$.  
Now straightforward applications of property (1) and (3) give the result.
%
%
%

\noindent{$\circ$ \bf{disjunction}}:
In case $\phi$ is a disjunction,  $\phi_1\vee\phi_2$ say, 
$\tupel{\csp ;\theta}\mysem{\phi} = 
\tupel{\csp ;\theta}\mysem{\phi_1}\cup\tupel{\csp ;\theta}\mysem{\phi_2}$.
The induction hypotheses apply immediately.
%

\noindent{$\circ$ \bf{conjunction}}:
In case $\phi$ is a conjunction, $\phi_1\wedge\phi_2$ say, 
$\tupel{\csp ''; \theta ''}\in\tupel{\csp ;\theta}\mysem{\phi} $ iff
$\tupel{\csp ''; \theta ''}\in\tupel{\csp ';\theta '}\mysem{\phi_2} $  for
some
$\tupel{\csp '; \theta '}\in\tupel{\csp ;\theta }\mysem{\phi_1} $.
Part 1. of the theorem is a straightforward consequence of the 
induction hypothesis and persistence. For part 2 we add some details.
We have: if $\tupel{\csp ';\theta '}\in\tupel{\csp ;\theta}\mysem{\phi_1}$ 
is consistent, 
then $\tupel{\csp ';\theta '}\mysem{\phi_2}$ only contains inconsistent states.
From this we may conclude by induction hypothesis that for each 
$\tupel{\csp ';\theta '}\in\mycons (\tupel{\csp ;\theta}\mysem{\phi_1})$
\begin{equation}
  \label{eq:otimes}
\tupel{\csp ' ;\theta '}\models_{\cal J}\neg \phi_2.  
\end{equation}
Now assume that for some\footnote{Notation for assignment of values as
in the lemma.} $[\overline{b},\overline{c},\overline{d},
\overline{e}]$,
$\models_{\cal J} (\csp\theta\wedge\phi_1\theta\wedge\phi_2\theta)[b,c,d,e]$
and that we have a 
$\tupel{\csp ';\theta '}\in\mycons (\tupel{\csp ;\theta}\mysem{\phi_1})$.
Then persistence (2) tells us that the consistency is preserved, i.e.,
there are 
$[\overline{b'},\overline{c'},\overline{d'},\overline{e'}]$ 
such that
$\models_{\cal J}
(\csp '\theta '\wedge\phi_1\theta '\wedge\phi_2\theta ')
[\overline{b'},\overline{c'},\overline{d'},\overline{e'}]$. 
But this contradicts the statement (\ref{eq:otimes}). So, for no 
$[\overline{b},\overline{c},\overline{d},\overline{e}]$, 
$\models_{\cal J} (\csp\theta\wedge\phi_1\theta\wedge\phi_2\theta)
[\overline{b},\overline{c},\overline{d},\overline{e}]$, 
which is as required.

\noindent{$\circ$ \bf{negation}}:
In case of a negation $\neg\phi$, there are three situations to consider
\begin{itemize}
\item $\tupel{\csp ';\theta '}\in\infer\tupel{\csp ;\theta}$ and 
$\mycons^+ (\tupel{\csp ;\theta}\mysem{\phi})=\emptyset$.
Now case (1) of the theorem follows from the induction hypothesis for 
case (2) and equivalence condition (1). Case (2) follows from 
conditions (1) and (3).
\item $\tupel{\csp ;\theta}\mysem{\neg\phi}=\emptyset$ and there is some
$\tupel{\csp ';\theta '}\in\mycons (\tupel{\csp ;\theta}\mysem{\phi})$,
which is equivalent  to $\tupel{\csp ;\theta}$. 
Now case (1) is satisfied trivially and case (2) follows from the induction
hypothesis for (1) and condition (1) on $\infer$.
\item $\tupel{\csp ;\theta}\mysem{\neg\phi}=
\infer\tupel{\csp ,{\neg\phi};\theta}$.
Let $\tupel{\csp ';\theta '}\in
\infer\tupel{\csp ,{\neg\phi};\theta}$ be given.
Now case (1) follows from condition (1) and case two relies on 
conditions (1) and (3).
\end{itemize}

\noindent{$\circ$ \bf{existential quantification}}:
In case of an existential quantification $\exists x\;\phi$, we have to consider
$\tupel{\csp '';\theta ''}\in 
\infer\;\mbox{\sc drop}_u\tupel{\csp ';\eta}$,
\noindent for $\tupel{\csp ';\eta}\in \mycons (
\tupel{\csp ;\theta}\mysem{\phi\verz{x/u}})$ (some fresh $u$). 
Call $\mbox{\it DROP}_{u,{\bf y}}(\eta )=\theta '$.
Below we use a crucial fact about first order logic:
if $x$ is not free in $\chi$, then
$\models\; \forall x\; (\psi\rightarrow\chi ) \;\leftrightarrow\;
((\exists x\;\psi)\rightarrow \chi )$. 

\begin{enumerate}
\item 
By induction hypothesis $\tupel{\csp ';\eta}\models_{\cal J}\phi\verz{x/u}$.
By first order logic

$(\csp '-\csp'(u,{\bf y}))\eta\models_{\cal J}(\csp '(u,{\bf y})\;\rightarrow\;
\phi\verz{x/u})\eta$.
From this we conclude that
$(\csp '-\csp'(u,{\bf y}))\eta\models_{\cal J}(\csp '(u,{\bf y})\;\rightarrow\;
\exists x\; \phi)\eta$.
$(\csp '-\csp '(u,{\bf y}))$ does not contain $u$ or ${\bf y}$, so:
$(\csp '-\csp'(u,{\bf y}))\theta '\models_{\cal J}(\csp '(u,{\bf y})\wedge
u=u\eta\wedge{\bf y}={\bf y}\eta\;\rightarrow\;
\exists x\; \phi)\theta '$.
As $u$ does not occur in $(\exists x\;\phi)\theta '$,
we can apply the crucial fact to get

$(\csp '-\csp'(u,{\bf y}))\theta '\;\wedge\;\exists u\; 
(\csp '(u,{\bf y})\;\wedge\;
u=u\eta\;\wedge\;{\bf y}={\bf y}\eta)\theta '\models_{\cal J}\;
(\exists x\; \phi)\theta '$.
Now we can make $\theta '$ more specific by re-instantiating the values
${\bf y}\eta$ for the variables in ${\bf y}$. 
This suffices (by (1)).
\item In this case there is no fresh $u$ which produces a 
        $\tupel{\csp ';\eta}\in\mycons (
        \tupel{\csp ;\theta}\mysem{\phi\verz{x/u}})$.
        The induction hypothesis then gives
        $\tupel{\csp ;\theta}\models_{\cal J}
        (\neg\phi\verz{x/u})$ (for all fresh $u$), from which
        $\tupel{\csp ;\theta}\models_{\cal J}(\exists x\;\phi)$ (as $u$ is fresh
        w.r.t. $\theta$). \hfill{$\Box$}
\end{enumerate} 


\begin{thebibliography}{10}

\bibitem{AS99a}
K.~R. Apt and A.~Schaerf.
\newblock The {Alma} project, or how first-order logic can help us in
  imperative programming.
\newblock In E.-R. Olderog and B.~Steffen, editors, {\em Correct System
  Design}, Lecture Notes in Computer Science 1710, pages 89--113, 1999.

\bibitem{Apt00}
K.R. Apt.
\newblock A denotational semantics for first-order logic.
\newblock In {\em Proc. of the computational logic conference (CL2000)},
  Lecture Notes in Artificial Intelligence 1861, pages 53--69. Springer Verlag,
  2000.

\bibitem{AB99}
K.R. Apt and M.A. Bezem.
\newblock Formulas as programs.
\newblock In K.R. Apt, V.W. Marek, M.~Truszcy\'nski, and D.S. Warren, editors,
  {\em The Logic Programming Paradigm: A 25 Year Perspective}, pages 75--107,
  1999.

\bibitem{ABPS98a}
K.R. Apt, J.~Brunekreef, V.~Partington, and A.~Schaerf.
\newblock {\sf Alma-0}: An imperative language that supports declarative
  programming.
\newblock {\em ACM Toplas}, 20(5):1014--1066, 1998.

\bibitem{dBGMP97}
F.S.~De Boer, M.~Gabbrielli, E.~Marchiori, and C.~Palamidessi.
\newblock Proving concurrent constraint programs correct.
\newblock In {\em ACM Transactions on Programming Languages and Systems},
  volume 19(5), pages 685--725, 1997.

\bibitem{Col84}
A.~Colmerauer.
\newblock Equations and inequations on finite and infinite trees.
\newblock In John Lloyd, editor, {\em Proc. of International Conference of
  Fifth Generation Computer Systems (FGCS'84)}, pages 85--99. OHMSHA Ltd. Tokyo
  and North-Holland, 1984.

\bibitem{dBdPP95}
F.S. de~Boer, A.~Di Pierro, and C.~Palamidessi.
\newblock Nondeterminism and infinite computations in constraint programming.
\newblock {\em Theoretical Computer Science}, 151(1):37--78, 1995.

\bibitem{FRS01}
F.~Fages, P.~Ruet, and S.~Soliman.
\newblock Linear concurrent constraint programming: Operational and phase
  semantics.
\newblock {\em Information and Computation}, 165(1):14--41, 2001.

\bibitem{jaffar-survey}
J.~Jaffar and J.M. Maher.
\newblock Constraint logic programming: a survey.
\newblock {\em Journal of Logic Programming}, 19/20, 1994.

\bibitem{JMMS98}
J.~Jaffar, J.M. Maher, K.~Marriott, and P.~Stuckey.
\newblock The semantics of constraint logic programs.
\newblock {\em Journal of Logic Programming}, 37(1):1--46, 1998.

\bibitem{LT84}
J.W. Lloyd and R.W. Topor.
\newblock Making {Prolog} more expressive.
\newblock {\em Journal of Logic Programming}, 1:225--240, 1984.

\bibitem{dBPdP97}
C.~Palamidessi, F.S. de~Boer, and A.~Di Pierro.
\newblock An algebraic perspective of constraint logic programming.
\newblock {\em Journal of Logic and Computation}, 7, 1997.

\bibitem{saraswat-semantic}
V.~A. Saraswat, M.~Rinard, and P.~Panangaden.
\newblock Semantic foundations of concurrent constraint programming.
\newblock In {\em Conference Record of the Eighteenth Annual {ACM} Symposium on
  Principles of Programming Languages}, pages 333--352, Orlando, Florida, 1991.

\bibitem{saraswat-ccp-93}
Vijay Saraswat.
\newblock {\em Concurrent Constraint Programming}.
\newblock MIT Press, 1993.

\end{thebibliography}
\end{document}